\UseRawInputEncoding
\documentclass[
 reprint,
 superscriptaddress,
 amsmath,
 amssymb,
 longbibliography,
 aps,
 prb,
 twocolumn,
 showpacs
]{revtex4-1}
\usepackage{graphicx}% Include figure files
\usepackage{dcolumn}% Align table columns on decimal point
\usepackage{bm}
\usepackage{epstopdf}
\usepackage[colorlinks,linkcolor=blue,citecolor=blue]{hyperref}
\usepackage{algorithm}
\usepackage{algorithmicx}
\usepackage{algpseudocode}
\usepackage{multirow}
\usepackage{booktabs}
\usepackage{changes}
\usepackage{CJKutf8}
\begin{document}

\title{Observation of size-dependent boundary effects in non-Hermitian electric circuits}
\author{Luhong Su\begin{CJK*}{UTF8}{gbsn}(苏鹭红)\end{CJK*}}
\affiliation{Beijing National Laboratory for Condensed Matter Physics, Institute of Physics, Chinese Academy of Sciences, Beijing 100190, China}
\affiliation{School of Physical Sciences, University of Chinese Academy of Sciences, Beijing 100049, China}
\author{Cui-Xian Guo\begin{CJK*}{UTF8}{gbsn}(郭翠仙)\end{CJK*}}
\thanks{Corresponding author: cxguo@iphy.ac.cn}
\affiliation{Beijing National Laboratory for Condensed Matter Physics, Institute of Physics, Chinese Academy of Sciences, Beijing 100190, China}
\author{Yongliang Wang\begin{CJK*}{UTF8}{gbsn}(王永良)\end{CJK*}}
\affiliation{CAS Center for Excellence in Superconducting Electronics, Shanghai Institute of Microsystem and Information Technology, Chinese Academy of Sciences, Shanghai 200050, China}
\author{Li Li\begin{CJK*}{UTF8}{gbsn}(李力)\end{CJK*}}
\affiliation{Beijing National Laboratory for Condensed Matter Physics, Institute of Physics, Chinese Academy of Sciences, Beijing 100190, China}
\affiliation{School of Physical Sciences, University of Chinese Academy of Sciences, Beijing 100049, China}
\author{Xinhui Ruan\begin{CJK*}{UTF8}{gbsn}(阮馨慧)\end{CJK*}}
\affiliation{Beijing National Laboratory for Condensed Matter Physics, Institute of Physics, Chinese Academy of Sciences, Beijing 100190, China}
\affiliation{Department of Automation, Tsinghua University, Beijing 100084, P. R. China}
\author{Yanjing Du\begin{CJK*}{UTF8}{gbsn}(杜燕京)\end{CJK*}}
\affiliation{Beijing National Laboratory for Condensed Matter Physics, Institute of Physics, Chinese Academy of Sciences, Beijing 100190, China}
\affiliation{China University of Geosciences, Beijing 100083, China}
\author{Shu Chen\begin{CJK*}{UTF8}{gbsn}(陈澍)\end{CJK*}}
\affiliation{Beijing National Laboratory for Condensed Matter Physics, Institute of Physics, Chinese Academy of Sciences, Beijing 100190, China}
\affiliation{CAS Center for Excellence in Topological Quantum Computation and School of Physical Sciences, University of Chinese Academy of Sciences, Beijing 100049, China}
\affiliation{The Yangtze River Delta Physics Research Center, Liyang, Jiangsu 213300, China}
\author{Dongning Zheng\begin{CJK*}{UTF8}{gbsn}(郑东宁)\end{CJK*}}
\thanks{Corresponding author: dzheng@iphy.ac.cn}
\affiliation{Beijing National Laboratory for Condensed Matter Physics, Institute of Physics, Chinese Academy of Sciences, Beijing 100190, China}
\affiliation{CAS Center for Excellence in Topological Quantum Computation and School of Physical Sciences, University of Chinese Academy of Sciences, Beijing 100049, China}
\affiliation{Songshan Lake Materials Laboratory, Dongguan, Guangdong 523808, China}

\begin{abstract}
The non-Hermitian systems with the non-Hermitian skin effect\,(NHSE) are very sensitive to the imposed boundary conditions and lattice sizes, which lead to size-dependent non-Hermitian skin effects. Here, we report the experimental observation of NHSE with different boundary conditions and different lattice sizes in the unidirectional hopping model based on a circuit platform. The circuit admittance spectra and corresponding eigenstates are very sensitive to the presence of the boundary. Meanwhile, our experimental results show how the lattice sizes and boundary terms together affect the strength of NHSE. Therefore, our electric circuit provides a good platform to observe size-dependent boundary effects in non-Hermitian systems.
\par\textbf{Keywords:} non-Hermitian, size-dependent boundary effects, circuit
\par\textbf{PACS:} 84.30-r, 03.65.vf,03.65.Fd, 03.65.Yz
\end{abstract}
\maketitle

\section{Introduction}
Over the past two decades, non-Hermitian systems have been widely studied due to their peculiar properties.
Contrast to Hermitian systems, the boundary condition plays an important role in non-Hermitian systems. The most fascinating phenomenon is the non-Hermitian skin effect (NHSE)\cite{PhysRevLett.121.086803}, which describes the accumulation of most of bulk states to the boundary under the open boundary condition\,(OBC), and it has been extensively investigated in various systems\cite{PhysRevLett.121.086803,PhysRevLett.116.133903, PhysRevX.8.031079, PhysRevResearch.1.023013,PhysRevLett.123.206404, PhysRevLett.123.246801, jiang2019interplay, PhysRevB.99.201103, PhysRevB.99.081103, PhysRevLett.125.226402, PhysRevLett.124.056802, PhysRevA.102.032203,PhysRevLett.125.126402,PhysRevLett.124.086801,PhysRevB.101.045415,PhysRevB.102.041119,PhysRevB.103.L041404,YUCE2021127484}. Systems with NHSE typically reflect the importance of boundary conditions, because their energy spectra and eigenstates under OBC are significantly different from those under periodic boundary condition\,(PBC). Recently, size-dependent NHSE has also attracted attention in some coupled non-Hermitian chains\cite{liu2020helical,li2020critical,PhysRevLett.127.116801} and non-reciprocal chains with impurity\cite{Li2021Communi,Lo2019}. And size-dependent NHSE has been explained from the perspective of exact solution \cite{PhysRevLett.127.116801}. In general, for a non-Hermitian system with NHSE under OBC, we expect a size-dependent NHSE for this system under generalized boundary conditions. Therefore, both the boundary conditions and the lattice sizes have a great impact on the non-Hermitian systems.

The progress of non-Hermitian theory has further promoted the observation of a variety of novel phenomena on various experimental platforms\cite{li2019observation,wang2019arbitrary, wu2019observation, CPAEP2021, QST2019, PhysRevResearch.2.013280, weidemann2020topological, XiaoL2020,qi2020robust, li2020topological, Ghatak29561,NRR,STP}. Due to the flexibility and simplicity in design and operation, electric circuits have become a powerful platform to realize non-Hermitian models and to explore novel phenomena\cite{liu2021non, GBBC, BSEC, hofmann2020reciprocal, 2019Non,TEC2019, TEC2015, TCcorner, BerryEC, PhysRevX.5.021031}, including spontaneous breaking of PT symmetry\cite{FYWu_2004, Schindler_2012}, non-reciprocal Su-Schrieffer-Heeger model\cite{GBBC,PhysRevB.103.014302}, non-Hermitian topological phases\cite{GBBC}.
However, the size-dependent boundary effect have rarely been realized and investigated in experiments.

In this paper, we demonstrate the size-dependent boundary effects of a unidirectional hopping model from the perspective of analytical results. Then we use $LC$ circuits combined with voltage followers\,(VF) to provide a useful platform for observing the size-dependent boundary effects. The key non-Hermitian component in our circuit is the VF, which implements unidirectional hopping. Through measuring the admittance spectra and eigenstates of this electric circuit, we confirm that the system is sensitive to  boundary conditions. Moreover, we also observe the interplay of boundary conditions and lattice sizes on the strength of NHSE. The circuit exhibits a size-dependent boundary sensitivity of NHSE in our experiment.

\section{The unidirectional hopping model}

\begin{figure*}[tbp]
\includegraphics[width=1\textwidth]{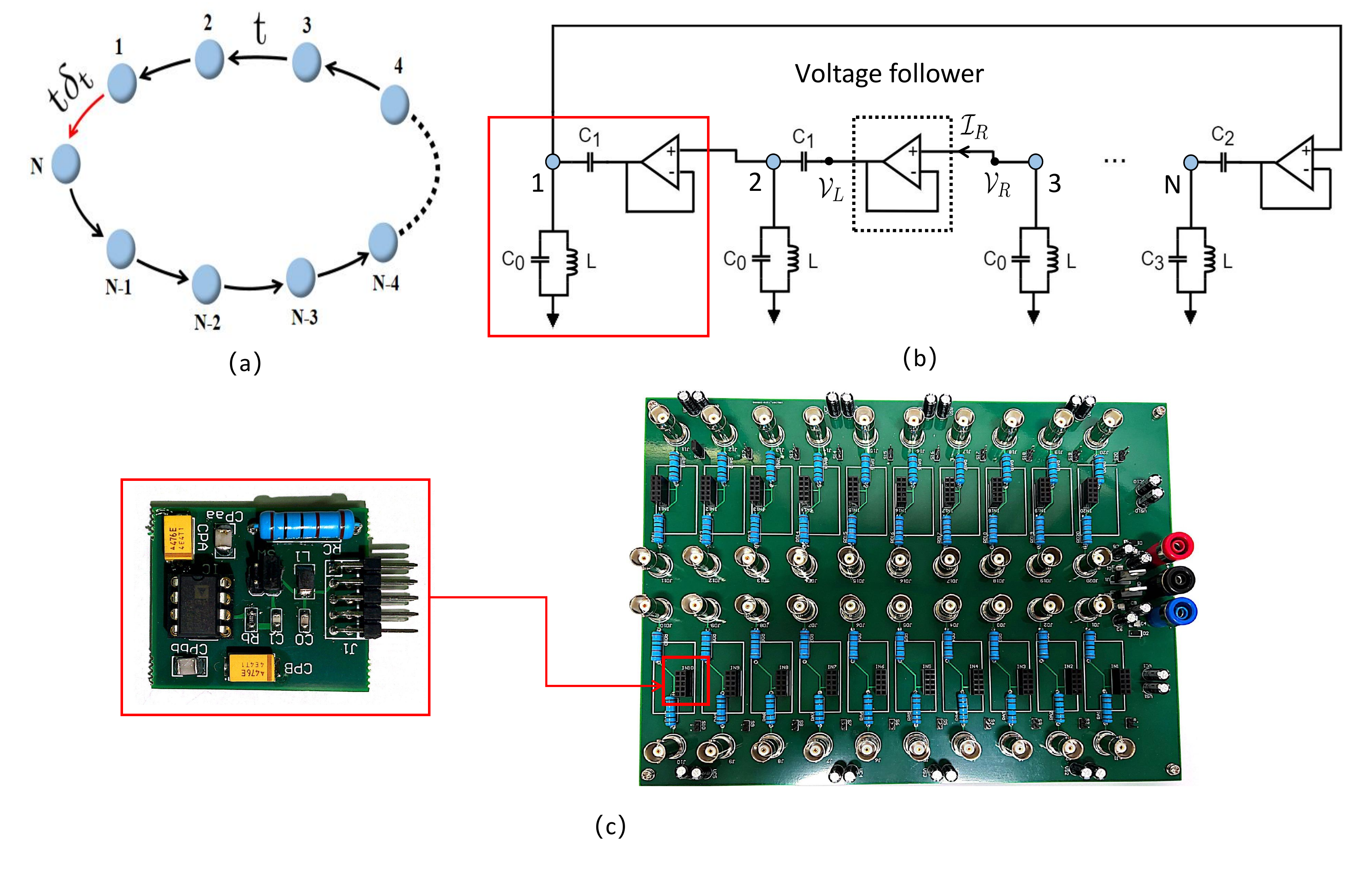}
\caption{(a) Schematic diagram of the unidirectional hopping model. (b) Schematic diagram of electrical circuit for realizing the unidirectional hopping model. (c) The printed circuit board\,(PCB) for realizing the unidirectional hopping model. In the right panel: the overall structure of the circuit board. In the left panel: circuit unit with 10 antennae needs to be plugged into the right circuit board. The circuit unit represents the unit inside the red box in (b).}
\label{fig1}
\end{figure*}

We introduce a unidirectional hopping model, which is a minimal non-Hermitian model with size-dependent boundary effects. The tight-binding Hamiltonian can be described as
\begin{equation}\label{H}
  H=\sum_{n=1}^{N-1}t|n\rangle\langle n+1|+t\delta_t |N\rangle\langle 1|,
\end{equation}
where $N$ is the number of lattice sites, $t$ denotes the unidirectional hopping amplitude, and $\delta_t$ determines the boundary condition. When $\delta_t=0$ the system is under OBC, while $\delta_t=1$ corresponds to PBC. The unidirectional hopping model is schematically displayed in Fig.\,\ref{fig1}(a).

The corresponding one-particle eigenvalue equation can be written as $H|\Psi\rangle=E|\Psi\rangle$, where $|\Psi\rangle=\sum_{n}\psi _{n}|n\rangle~(n=1,\cdots,N)$. This eigenvalue equation could be decomposed into a series of equations that describe the bulk and boundaries of the system. Thus, the bulk equation can be expressed as
\begin{equation}\label{H1BE}
-E\psi _{n}+t\psi _{n+1}=0\text{ \ \ }(n=1,2,\cdots ,N-1),
\end{equation}
and the boundary equation is given by
\begin{equation}\label{H1BdE}
t\delta_t \psi _{1}-E\psi _{N}=0.
\end{equation}
In consideration of the spatial translational nature of bulk equations,
we assume that the wave function $|\Psi\rangle$ has the following form
\begin{equation}\label{H1wv}
|\Psi\rangle =(\psi _{1},\psi _{2},\cdots ,\psi _{N})^{T}=(1,z,z^{2},\cdots ,z^{N-1})^{T},
\end{equation}
where $\psi_{n}=z^{n-1}(n=1,\cdots ,N)$, and $z$ is a complex function.

After inserting Eq.\,(\ref{H1wv}) into Eq.\,(\ref{H1BE}), the eigenvalue can be obtained as
\begin{equation}\label{H1E}
E=tz.
\end{equation}

Then we insert the ansatz of wave function Eq.\,(\ref{H1wv}) and expression of
eigenvalue Eq.\,(\ref{H1E}) into boundary equation Eq.\,(\ref{H1BdE}), and we have
\begin{equation}
\delta_t -z^{N}=0.
\end{equation}
The above equation gives rise to the solution of $z$:
\begin{equation}
z=\sqrt[N]{\delta_t}e^{i\theta }\ \ \left(
\theta =\frac{2m\pi }{N}\text{ \ \ }\left( m=1,2,\cdots ,N\right)
\right).
\end{equation}

Accordingly, the energy spectrum can be re-expressed as
\begin{equation}
E=t\sqrt[N]{\delta_t}e^{i\theta },
\end{equation}
and eigen wavefunctions can be re-written as
\begin{equation}
|\Psi\rangle =\bigg(1,\sqrt[N]{\delta_t}e^{i\theta },(\sqrt[N]{\delta_t}e^{i\theta })^{2},\cdots ,(\sqrt[N]{\delta_t}e^{i\theta })^{N-1}\bigg)^{T}.
\end{equation}

In non-Hermitian systems, energy-degenerate points with states coalescence are called exceptional points\,(EPs). When $\delta_t=0$, all eigenvalues are degenerate at $E=0$, and all eigenstates fuse into a highly localized state $(1,0,\cdots,0)^T$ independent of lattice size $N$. Therefore, $\delta_t=0$ corresponds to $N$th-order EP. It is obvious that the NHSE occurs in this case since all bulk states are localized at the left edge. When $\delta_t=1$\,(PBC), we have $|z|=1$, which indicated that all eigenstates are extended states independent of lattice size $N$. When $\delta_t\neq 0$ and $\delta_t\neq 1$, there are NHSE due to $|z|=|\sqrt[N]{\delta_t}|\neq 1$, and the strength of NHSE is related to boundary coupling $\delta_t$ and lattice size $N$. We can see that the more $|z|$ deviates from 1, the greater the strength of NHSE becomes. More specifically, the strength of NHSE increases with the decrease of $\delta_t$ and the decrease of $N$ for the case of $0<|z|<1$, which is the case this paper focused on from $0<\delta_t<1$.

\section{Realization of unidirectional electric circuit}

In this section, we employ circuit components including capacitors, inductors, and VFs to realize a unidirectional electric circuit lattice, as illustrated in Fig.\,\ref{fig1}(b). In this circuit, the value of $C_0$\,(and $C_1$) is the same for all sites, except for the last site. In the last site, the value of $C_2$ can be adjusted in order to vary the boundary conditions. In other words, we may set $C_2=0$ to achieve OBC and set $C_2=C_1$ to show the PBC. Additionally, as discussed in the follows,  it is required that $C_1+C_0=C_2+C_3$.

In general, capacitors and inductors are used as reciprocal devices, and they satisfy
\begin{equation}
C\frac{d\mathcal{V}}{dt}=\mathcal{I},~~~L\frac{d\mathcal{I}}{dt}=\mathcal{V}.
\end{equation}
The key unidirectional hopping effect is implemented with the help of VF. According to the characteristics of high input impedance and low output impedance of VF, we get the current and voltage relation as
\begin{equation}
\mathcal{V}_{R}=\mathcal{V}_{L},~~~~\mathcal{I}_{L}=0,
\end{equation}
where $\mathcal{V}_{R},~\mathcal{V}_{L}$, and $\mathcal{I}_{L}$ are shown in Fig.\,\ref{fig1}(b).

Here, we consider the case of applying alternating current\,(AC) source with frequency $\omega=2\pi f$. The  current and the corresponding voltage response may have the following form
\begin{equation}
\mathcal{I}(t)=Ie^{i\omega t},~~~\mathcal{V}(t)=Ve^{i\omega t}.
\end{equation}
According to the Kirchhoff's law, the circuit system is fundamentally described by\cite{liQH}
\begin{equation}
I_{m}=\sum_ng_{mn}(V_m-V_n)+w_mV_m
\end{equation}
where $I_{m}$ and $V_m$ are the input current and voltage at node $m$, respectively. $g_{mn}$ is the corresponding conductance when the current flowing out of node $m$ towards node $n$, and $w_{m}$ is the conductance when the current flowing out of node $m$ towards the ground. In a compact form, the circuit lattice can be described by the admittance matrix (or the circuit Laplacian) as
\begin{equation}
\mathbf{I}(\omega)=J(\omega)\mathbf{V}(\omega)
\end{equation}
where the current and voltage vectors are defined as $\mathbf{I}=(I_1,I_{2},\cdots)$ and $\mathbf{V}=(V_1,V_{2},\cdots)$, respectively, and $J$ is the admittance matrix (or the circuit Laplacian).

Therefore, the $N\times N$ admittance matrix for circuit lattice shown in Fig.\,\ref{fig1}(b) can be obtained as
\begin{equation}
\begin{split}\label{J}
  J(\omega)&=-i\omega\left(
      \begin{array}{ccccccc}
        \mu_1 & C_1 & 0 & 0 & \cdots & 0 & 0 \\
        0 & \mu_1 & C_1 & 0 & \cdots & 0 & 0 \\
        0 & 0 & \mu_1 & C_1 & \cdots & 0 & 0 \\
        0 & 0 & 0 & \mu_1 & \cdots & 0 & 0 \\
        \cdots & \cdots & \cdots & \cdots & \cdots & \cdots & \cdots \\
        0 & 0 & 0 & 0 & \cdots &  & C_1 \\
        C_2 & 0 & 0 & 0 & \cdots & 0 & \mu_2 \\
      \end{array}
    \right),
\end{split}
\end{equation}
where $\mu_1=\frac{1}{\omega^2 L}-(C_1+C_0)$, $\mu_2=\frac{1}{\omega^2 L}-(C_2+C_3)$. When the relationship between capacitors $C_2+C_3=C_1+C_0$ is satisfied, the equation above becomes
\begin{equation}
\begin{split}\label{JJJ}
  J(\omega)&=-i\omega\left(
      \begin{array}{ccccccc}
        0 & C_1 & 0 & 0 & \cdots & 0 & 0 \\
        0 & 0 & C_1 & 0 & \cdots & 0 & 0 \\
        0 & 0 & 0 & C_1 & \cdots & 0 & 0 \\
        0 & 0 & 0 & 0 & \cdots & 0 & 0 \\
        \cdots & \cdots & \cdots & \cdots & \cdots & \cdots & \cdots \\
        0 & 0 & 0 & 0 & \cdots & 0 & C_1 \\
        C_2 & 0 & 0 & 0 & \cdots & 0 & 0 \\
      \end{array}
    \right)-i\omega\mu(\omega) I_{N\times N},\\
    &=\widetilde{J}(\omega)-i\omega\mu(\omega) I_{N\times N},
\end{split}
\end{equation}
where $\mu(\omega)=\mu_1=\mu_2=\frac{1}{\omega^2 L}-(C_1+C_0)$, $I_{N\times N}$ denotes a $N\times N$ identity matrix.
By comparing Eq.\,(\ref{JJJ}) and Eq.\,(\ref{H}), we find that there is a mapping between $H$ and $\widetilde{J}$ with $t\sim (-i\omega C_1)$. In this paper, we call both $J$ and $\widetilde{J}$ the admittance matrix because they differ by only one identity matrix term, and they share the same eigenvetors. Therefore, this electric circuit is suitable for investigating the size-dependent boundary effects of the unidirectional hopping model.\\

In the experiment, a circuit consisting of multiple units arranged in a one-dimensional chain is realized on printed circuit board\,(PCB), as illustrated in Fig.\,\ref{fig1}(c).
In the design of PCBs, we have used the circuit boards with four layers, one for signal transmission, two for the supply of direct current voltage for the operational amplifier\,(OpAmp), and one for grounding. For each site of the lattice, the unit inside the red box in Fig.\,\ref{fig1}(b) consists of a LC circuit and a capacitor and a VF. The capacitor provides the coupling between the adjacent nodes and the VF ensures the unidirectional signal transmission.
To conveniently adjust boundary coupling and lattice size, we use the circuit unit with 10 antennae shown in the left panel of Fig.\,\ref{fig1}(c) to realizing the unit of model, and then insert it into the main circuit board described in the right panel of Fig.\,\ref{fig1}(c).
In order to reduce the crosstalk between two adjacent inductors, we purposely increase the distance between two lattice sites to about 4 cm.
Because the VF is an active circuit element, it pumps energy into the system and might cause self-oscillation.
 The caution measures are taken to prevent these self-oscillation by including some damping resistors and shunt capacitors in the circuits, which are not displayed in Fig.\,\ref{fig1}(b). In the implementation of the VF, we choose OpAmp OP07, and its gain bandwidth product is 600\,kHz. In order to maintain the stability of unity-gain stability for VF, we adjust the self oscillation frequency of the circuit to about 700\,kHz, so the OP07 will not generate gain for self oscillation frequency.

\section{Observation of NHSE with different boundary conditions}
\begin{figure}[tbp]
\includegraphics[width=0.48\textwidth]{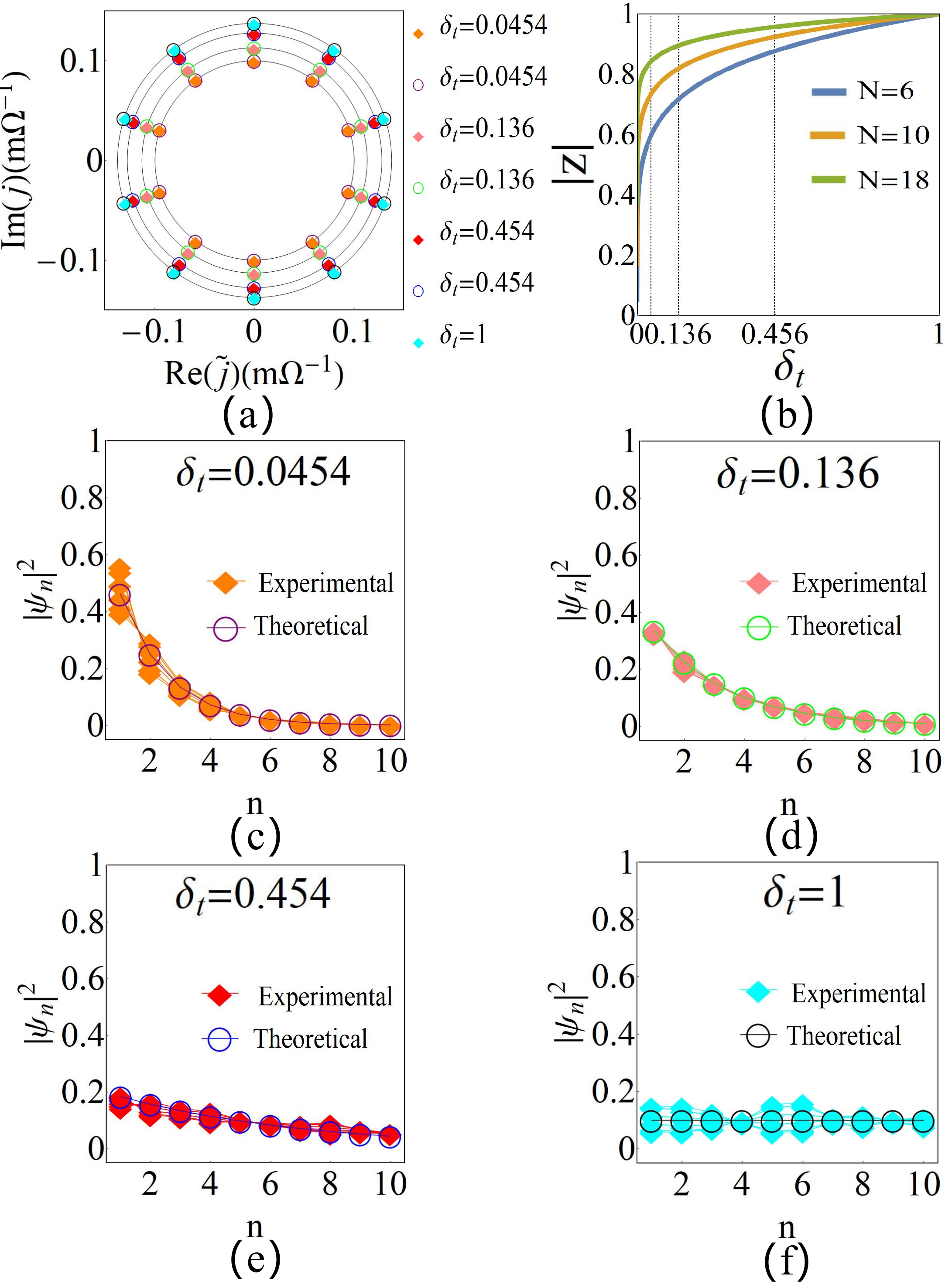}
\caption{(a) Eigenvalues of the admittance matrix $\widetilde{J}(\omega)$ with different boundary coupling $\delta_t$. The hollow circles represent theoretical data, while the solid diamonds represent experimental data. (b) $|z|$ as a function of $\delta_t$ for different $N$ from theory, which reflect the strength of the NHSE. (c-f)
The profile of all eigenstates with $\delta_t=0.0454$, $\delta_t=0.136$, $\delta_t=0.454$ and $\delta_t=1$, respectively. Common parameters: $N=10, f=100$\,kHz.}
\label{fig2}
\end{figure}

In this section, we provide a more detailed description about the admittance band measurement and show the results. As discussed in the previous section, the eigenvalues $\widetilde{j}(\omega)$ and eigenvectors $\psi$ of the admittance matrix $\widetilde{J}(\omega)$ behave similarly to those of $H$. In practice, it is more convenient to perform impedance measurements than to measure admittance. Thus, in experimental measurements, we measure the impedance matrix $G$ first and then perform matrix inversion to obtain the admittance matrix $\widetilde{J}(\omega)$\,(or $J(\omega)$). The impedance matrix $G$ is defined as
\begin{equation}
\mathbf{V}(\omega)=G(\omega)\mathbf{I}(\omega).
\end{equation}
 If we apply an input AC current $I(f)$ at a specified node $n$ of the circuit, the corresponding elements of impedance matrix can be obtained as
\begin{equation}
G_{mn}=\frac{V_{m}^{n}}{I_{n}}=(J^{-1})_{mn},
\end{equation}
where $V_{m}^{n}$ represents the voltage response at node $m$ when the only input current is given by $I_n$ at node $n$. Then we can obtain the admittance matrix $J(\omega)$ according to the relation $J(\omega)=G^{-1}(\omega)$, and we have $\widetilde{J}(\omega)=J(\omega)+i\omega\mu(\omega) I_{N\times N}$.

In the experiment, the AC current is provided by an AC voltage source\,(NF Wave Factory1974) connected to the PCB through a resistance of $R=2k\Omega$, while a lock-in amplifier\,(Zurich Instruments UHF) is employed for the measurement of voltage response of the other nodes. We measure a steady state response, so that the voltage response of each node is proportional only to its associated impedance eigenvalue instead of being affected by dynamical damping. The measurement frequency is set as 100 kHz.

Here, we have built a circuit chain with $N =10$ to study the influence of boundary conditions. In this case, we set $C_0=10$\,nF, $C_1=220$\,nF, $L=220$\,$\mu$H, and the strength of boundary coupling is adjusted by changing $C_2$ and $C_3$. The parameters of the components used for the circuit implementation are shown in table\,\ref{table1}.

\begin{table}[htb]
\begin{center}
\caption{Parameters of inductances and capacitances corresponds to different boundary couplings.}
\label{table1}
\renewcommand{\arraystretch}{1.5}
\begin{tabular}{|c|c|c|c|c|c|}
\hline
\textbf{Inductance($\mu$H)}&\multicolumn{4}{c|}{\textbf{Capacitance(nF)}}& \textbf{Boundary coupling}\\
\hline
 $L$ &\textbf{$C_0$}&\textbf{$C_1$}&\textbf{$C_2$}&\textbf{$C_3$}&$\delta_t$ \\
\hline
\multirow{5}{*}[8pt]{220}&\multirow{5}{*}[8pt]{10} &\multirow{5}{*}[8pt]{220} &10 &220 & 0.0454\\
\cline{4-6}
& &  &30 &200 & 0.136\\
\cline{4-6}
 &  & &100&130 & 0.454\\
 \cline{4-6}
  &  & &220&10 &1 \\
\hline
\end{tabular}
\end{center}
\end{table}

After measuring the $G(\omega)$ matrices, the admittance matrices $J(\omega)$, the corresponding eigenvalue $j(\omega)$ and eigenfunction $\psi$ are obtained by numerical diagonalization. Further, for $\widetilde{J}(\omega)$, the eigenvalues can be obtained as $\widetilde{j}(\omega)=j(\omega)+i\omega \mu(\omega)$, while the eigenvetors of $\widetilde{J}(\omega)$ are the same as those of $J(\omega)$.\\

In Fig.\,\ref{fig2}, we present the eigenvalues and eigenstates of the admittance matrix $\widetilde{J}(\omega)$ with different boundary couplings $\delta_t$ including theoretical data and experimental data. From Fig.\,\ref{fig2}(a), we find that the absolute values of the eigenvalues increase as $\delta_t$ increases due to $|E|=|t||z|$. The behavior of $|z|$ as a function of $\delta_t$ for different $N$ are illustrated in Fig.\,\ref{fig2}(b).
As shown in Fig.\,\ref{fig2}(c), when $\delta_t=0.0454$, NHSE is the most obvious. As $\delta_t$ increases from $\delta_t=0.0454$ to $\delta_t=1$, the strength of the skin effect gradually weakens, accompanied by the absolute values of the eigenvalues increases, just as predicted by the theory. The closer $|z|$ is to 1, the closer the eigenstates is to the extended states. $\delta_t=1$ corresponds to the PBC, at which eigenstates become extended states and $|z|=1$. So we have observed that the eigenstates change from NHSE to extended states when $\delta_t$ increases from $\delta_t=0.0454$ to $\delta_t=1$. The experimental data are in great agreement with the theoretical data. Therefore, this observation explicates the sensitivity of the admittance spectrum with respect to boundary conditions.

\begin{figure}[hbp]
\includegraphics[width=0.48\textwidth]{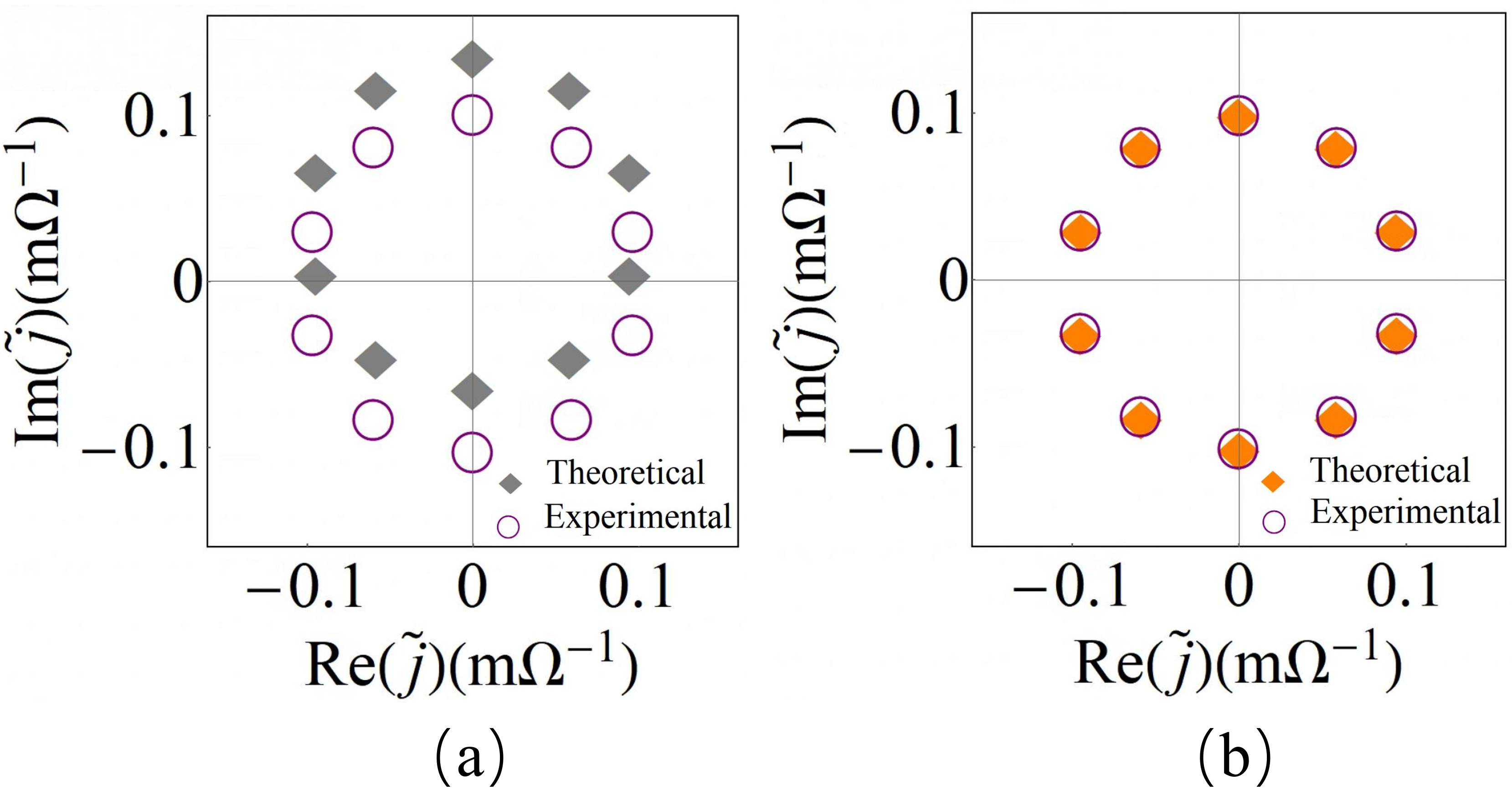}
\caption{Eigenvalues of the admittance matrix $\widetilde{J}(\omega)$ with $N=10,\,\delta_t=0.0454,\,f=100$\,kHz. The hollow circles represent theoretical data, while the solid diamonds represent experimental data.  (a) The results are obtained from original data. (b) The results are obtained after accounting for the device errors of $1.5\%$ to capacitances and $0.54\%$ to inductances.}
\label{fig2s}
\end{figure}

\begin{table}[htb]
\begin{center}
\caption{The specific correction data of the device errors related to the capacitances and inductances for the results shown in Fig.\,\ref{fig2} corresponding $N=10$ with different $\delta_t$.}
\label{table2}
\renewcommand{\arraystretch}{1.5}
\begin{tabular}{|c|c|c|c|c|}
\hline
 & $\delta_t=0.0454$  &  $\delta_t=0.136$   &     $\delta_t=0.454$  &   $\delta_t=1$  \\
\hline
$L$  &   $L+0.0054$   &   $L+0.0051$   &   $L+0.005$   &   $L+0.007$   \\
\hline
$C_i$  &   $C_i-0.015$   &   $C_i-0.002$   &   $C_i$   &   $C_i-0.005$  \\
\hline
\end{tabular}
\end{center}
\end{table}

There are two main factors affecting the measurement results, one is the circuit noise, the other is device errors. During these experiments, we have adopted a series of methods to obtain the satisfactory results shown in these figures. Due to the influence of the circuit noise on measurement results, it is important to improve the signal to noise ratio\,(SNR). In our experiments, the noise source mainly come from two parts, one is self-oscillation and the other is the thermal white noise. In order to reduce the effects of noise, we enhanced the signal of AC source on one hand, and we employed the lock-in amplifier to measure voltage response on the other hand.

It is inevitable that there are parameter errors in the manufacturing process of circuit devices.
The device errors are mainly caused by the fact that the actual values of these devices such as capacitances and inductances are not exactly equal to the nominal values. We use the following two steps to reduce the impact of device errors. Firstly, for these circuit units with the same nominal parameters, we carefully select these circuit units whose actual parameters are close to each other and insert them into the main circuit board. Secondly, we make uniform correction below $2\%$ for all inductances and capacitances in the original data.
In fact, when comparing the experimental and theoretical data of the eigenvalues $\widetilde{J}(\omega)$ in Fig.\,\ref{fig2}, we have considered the device errors below $2\%$ from capacitances and inductances, which is a reasonable correction. Here, we take the case of $N=10,\,\,\delta_t=0.0454$ as an example to display the correction of device errors in detail. As shown in Fig.\,\ref{fig2s}(a), we present the eigenvalues of the admittance matrix $\widetilde{J}(\omega)$ including theoretical data and experimental data from original data. However, there is a shift between the theoretical data and experimental data, thus it is difficult to compare. After accounting for the device errors of $1.5\%$ to all capacitance and $0.54\%$ to all inductances, the experimental data are in great agreement with the theoretical data as shown in Fig.\,\ref{fig2s}(b). In table\,\ref{table2}, we give the specific correction data of the capacitances and inductances for the results shown in Fig.\,\ref{fig2} corresponding $N=10$ with different $\delta_t$. In addition, in the process of obtained eigenstates $\psi$, the experimental data are obtained from the original data are agree with the behavior predicted by the theory. Therefore, the results we present are reliable.

\section{Observation of the size-dependent NHSE}
\begin{figure*}[tbp]
\includegraphics[width=1\textwidth]{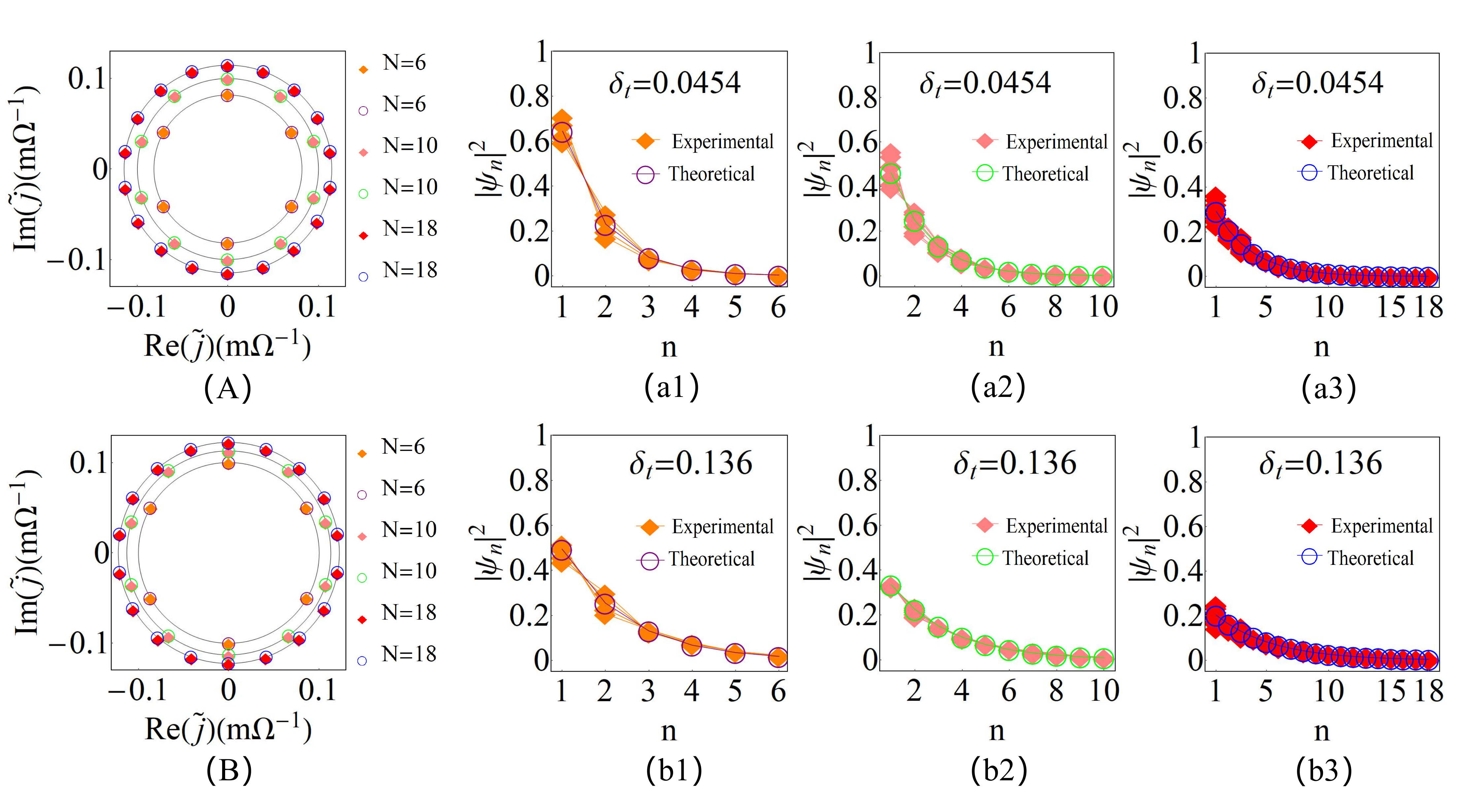}
\caption{(A) Eigenvalues of the admittance matrix $\widetilde{J}(\omega)$ with different lattice sites when $\delta_t=0.0454$. The hollow circles represent theoretical data, while the solid diamonds represent experimental data. (a1-a3)
The profile of all eigenstates with $N=6$, $N=10$, and $N=18$ when $\delta_t=0.0454$, respectively.
(B) Eigenvalues of the admittance matrix $\widetilde{J}(\omega)$ with different lattice sites when $\delta_t=0.136$. The hollow circles represent theoretical data, while the solid diamonds represent experimental data. (b1-b3)
The profile of all eigenstates with $N=6$, $N=10$, and $N=18$ when $\delta_t=0.136$, respectively.
Common parameter: $f=100$\,kHz.}
\label{fig3}
\end{figure*}

The circuit platform does not only show NHSE with different boundary conditions, but also can simulate the interplay of boundary hopping terms and lattice sizes. Using the same method, we have built circuit boards of $N=6,~10$, and $18$ lattice sites with $\delta_t=0.136$ and $0.0454$. The experimental and theoretical results are shown in Fig.\,\ref{fig3}. The hollow circles represent theoretical data, while the solid diamonds represent experimental data. Similarly, the results shown in Fig.\,\ref{fig3} are also obtained after considering the device errors of capacitances and inductances, and the specific correction data are given in tables\,\ref{table3} and  \ref{table4}.

In this case, we also set $C_0=10$\,nF, $C_1=220$\,nF,$L=220$\,$\mu$H. For the case of $\delta_t=0.0454$ with $C_2=10$\,nF, $C_3=220$\,nF, the absolute values of the eigenvalues increase as $N$ increases, as illustrated in Fig.\,\ref{fig3}(A). The data in Figs.\,\ref{fig3}(a1-c1) indicate that the strength of NHSE become decrease as $N$ increases. In Fig.\,\ref{fig3}(B) and Figs.\,\ref{fig3}(b1-b3), we also display the eigenvalues and eigenstates with different lattice sites for the case of $\delta_t=0.136$ with $C_2=30$\,nF, $C_3=200$\,nF. Both examples demonstrate the existence of size-dependent NHSE. Meanwhile, when boundary coupling $\delta_t$ deviates from 1, the size-dependent NHSE becomes more obvious. Comparing the data in Figs.\,\ref{fig3}(a1-a3) and Figs.\,\ref{fig3}(b1-b3), we find that the eigenstate for the case of $\delta_t=0.136$ tends to the extended state faster than those for the case of $\delta_t=0.0454$ with the increase of size. The experimental data are in great agreement with the theoretical data.\\

\begin{table}[htb]
\begin{center}
\caption{The specific correction data of the capacitances and inductances for the results shown in Fig.\,\ref{fig3} corresponding $\delta_t=0.0454$ with different $N$.}
\label{table3}
\renewcommand{\arraystretch}{1.5}
\begin{tabular}{|c|c|c|c|}
\hline
 & $N=6$  &  $N=10$   &     $N=18$   \\
\hline
$L$  &   $L+0.0048$   &   $L+0.0054$   &   $L+0.0048$     \\
\hline
$C_i$  &   $C_i-0.01$   &   $C_i-0.015$   &   $C_i-0.015$    \\
\hline
\end{tabular}
\end{center}
\end{table}

\begin{table}[htb]
\begin{center}
\caption{The specific correction data of the capacitances and inductances for the results shown in Fig.\,\ref{fig3} corresponding $\delta_t=0.136$ with different $N$.}
\label{table4}
\renewcommand{\arraystretch}{1.5}
\begin{tabular}{|c|c|c|c|}
\hline
 & $N=6$  &  $N=10$   &     $N=18$   \\
\hline
$L$  &   $L+0.0045$   &   $L+0.0051$   &   $L+0.0048$     \\
\hline
$C_i$  &   $C_i+0.01$   &   $C_i-0.002$   &   $C_i-0.01$    \\
\hline
\end{tabular}
\end{center}
\end{table}

\section{Conclusion}\label{s5}
\indent Using the simplest but prototypical non-Hermitian model\,(unidirectional hopping model), we present the size-dependent boundary effects by realizing the corresponding electric circuits. In general, there is NHSE when one-dimensional non-reciprocal chain is under OBC. Here, we have observed the NHSE even when the system has a tiny boundary perturbation $\delta_t~\rightarrow~0$. As $\delta_t$ ($0<|\delta_t|<1$) increases, the strength of NHSE gradually decrease. When $\delta_t=1$ (PBC), all bulk states become extended states as expected. Moreover, we confirmed a size-dependent NHSE in our circuit. For a fixed boundary coupling $\delta_t$, large size $N$ has an effect of repression on the strength of NHSE.
As our experimental results show, while the system may exhibit NHSE for a finite size $N$, the NHSE may disappears in the large size limit as $N\rightarrow\infty$ for a fixed $\delta_t$. Therefore, the behavior of eigenstates for our system in the thermodynamic limit is similar to those under PBC. Electric circuits can be used to explore more uncharted territory with an unprecedented degree of tunability and measurability. For instance, it also provides a potential scheme for further research of critical skin effect in the future.

\begin{acknowledgments}
\indent This work is supported by the State Key Development Program for Basic Research of China (Grant No. 2017YFA0304300), the Key-Area Research and Development Program of Guangdong Province, China (Grant No. 2020B0303030001), the National Natural Science Foundation of China (Grant No. T2121001) and the Strategic Priority Research Program of Chinese Academy of Sciences (Grant No. XDB28000000).
\end{acknowledgments}

\bibliography{ref}

\end{document}